# SG-PBFT: a Secure and Highly Efficient Blockchain PBFT Consensus Algorithm for Internet of Vehicles

Guangquan Xu, *Member, IEEE*, Yihua Liu, Jun Xing, Tao Luo\*, Yonghao Gu, Shaoying Liu, *Fellow, IEEE*, Xi Zheng\*, Athanasios V. Vasilakos, *Senior Member, IEEE*

*Abstract*—The Internet of Vehicles (IoV) is an application of the Internet of things (IoT). It faces two main security problems: (1) the central server of the IoV may not be powerful enough to support the centralized authentication of the rapidly increasing connected vehicles, (2) the IoV itself may not be robust enough to single-node attacks. To solve these problems, this paper proposes SG-PBFT: a secure and highly efficient PBFT consensus algorithm for Internet of Vehicles, which is based on a distributed blockchain structure. The distributed structure can reduce the pressure on the central server and decrease the risk of single-node attacks. The SG-PBFT consensus algorithm improves the traditional PBFT consensus algorithm by using a score grouping mechanism to achieve a higher consensus efficiency. The experimental result shows that our method can greatly improve the consensus efficiency and prevent single-node attacks. Specifically, when the number of consensus nodes reaches 1000, the consensus time of our algorithm is only about 27% of what is required for the state-of-the-art consensus algorithm (PBFT). Our proposed SG-PBFT is versatile and can be used in other application scenarios which require high consensus efficiency.

## I. INTRODUCTION

### A. Motivation

The Internet of Vehicles (IoV) has emerged as a new technology to support automatic, intelligent driving and to improve traffic services and operational efficiency [1][2]. IoV is usually applied in an open wireless network environ- ment whose topology changes rapidly [3] [4]. The traditional Internet of Vehicles is connected by a central server. Although this structure is simple and easy to control terminals, once the central server is attacked, the whole system may be paralyzed [5]. Moreover, as the scale of the network continues to expand, the burden on the central server is also increasing. This can easily lead to increased network delays and even network collapse.

The communication between vehicles in the IoV is the realization of information exchange and information sharing between a group of vehicles [6]. This information includes the vehicles status information such as the vehicles location and the driving speed, which can be used to determine road traffic conditions. Therefore, identity authentication is an important process for vehicles to communicate. When the vehicles communicate with each other, the malicious vehicles may attack the communication process. The vehicles and the base stations are also likely to be attacked [7] [8]. Furthermore, the malicious vehicles may send false messages, such as congestion in front of the road, or send false signals indicating that the road is clear [9]. This may cause the waste of travel time. If the base station is attacked, the attacker can tamper, add, and delete some data in the base station [10]. This situation may reveal the privacy information of some users [11]. Therefore, the reliability inspection of vehicles is the key to ensure the security of the IoV system.

To improve the security of the vehicles' identity authentica- tion, identity authentication framework [12] and authentication protocol [13] are proposed to improve the interactivity of com- munication between vehicles [14]. Then, the authentication methods based on blockchain technology were successfully proposed. Compared with the previous scheme, the method of using blockchain technology can more effectively ensure the users security. Therefore, the distributed architecture of blockchain technology is a better choice for the IoV system [15].

Guangquan Xu is with the Big Data School, Qingdao Huanghai University, Qingdao, China, 266427 and the Tianjin Key Laboratory of Advanced Net- working (TANK), College of Intelligence and Computing, Tianjin University, Tianjin, China, 300350. E-mail:(losin@tju.edu.cn).
Yihua Liu is with the International Engineering Institute, Tianjin University, Tianjin, China, 300350. E-mail:(13920878179@163.com).
Jun Xing is with the Big Data School, Qingdao Huanghai University, Qingdao, China, 266427. E-mail:(fumulan_cn@163.com).
Tao Luo is with the College of Intelligence and Computing, Tianjin University, Tianjin 300350, China Tianjin, E-mail:(luo tao@tju.edu.cn).
Yonghao Gu is with the school of Computer Science, Beijing University of Posts and Telecommunications, Beijing 100876 and the Beijing Key Laboratory of Intelligent Telecommunications Software and Multimedia, Beijing University of Posts and Telecommunications, Beijing 100876. E-mail:(guyonghao@bupt.edu.cn).
Shaoying Liu is with School of Informatics and Data Science, Hiroshima University, Japan. E-mail:(sliu@hiroshima-u.ac.jp).
Xi Zheng is with the Department of Computing, Macquarie University, Sydney, NSW 2109, Australia, E-mail:(james.zheng@mq.edu.au).
Athanasios V. Vasilakos is with the University of Technology Sydney, Australia, with the Fuzhou University, Fuzhou, China, and Lulea University of Technology, Lulea, Sweden, E-mail:(th.vasilakos@gmail.com).
\*Corresponding authors: Tao Luo, Xi Zheng

Since the distributed structure of blockchain is less prone to server bottlenecks and server paralysis, it is very appropriate to use the advantages of blockchain to solve the problems in the IoV system [16]. The consensus algorithm is the key method to ensure the consistency of blockchain. The Paxos algorithm has a highly fault-tolerant mechanism [17]. The Paxos protocol proposes that as long as (f + 1) of the (2f + 1) nodes in the system are non-faulty, the entire system is fault-tolerant and high data consistency can be guaranteed, where f is the total number of faulty nodes in the underlying network. It can also ensure the strong consistency of data, and there will be no problems such as data loss or conflict due to a few node downtime, hardware failure, message loss, network segmentation and other abnormalities. This largely guarantees the overall activity. However, in the Paxos algorithm, if two proposers propose two requests with increments numbers in turn, it may result in a dead-cycle of proposings to rejection that leaves the Paxos unable to get a final result, thereby inactivating the system. Another widely used blockchain consensus algorithm is PBFT. The basic principle of PBFT is based on a specific consensus protocol [18]. Everyone on the chain participates in the vote. If less than (N-1)/3 nodes against the request (N is the total number of the nodes), they have the right to publicize the information. The time to achieve consensus based on the voting in the algorithm, which we will call consensus time in this paper, is about 2 to 5 seconds, but it cannot meet the actual requirements of IoV.

### B. Main contributions

We propose a novel high efficient PBFT consensus algorithm for IoV in this paper. This scheme can ensure high security and low communication overhead of the Internet of Vehicles.

The main contributions of this paper are as follows:

1) We propose an optimized SG-PBFT consensus algo- rithm. Our algorithm is more efficient, has less commu- nication overhead, and has greater throughput than the original PBFT algorithm. Our algorithm is more suitable for the IoV system.

2) We propose a Vehicle - Service Provider – Roadside Unit architecture to ensure the orderliness and safety of vehicles.

3) We propose an identity authentication scheme based on blockchain. This scheme adopts blockchain technology and cryptography to ensure the system security.

4) We conduct the experiments to compare our solution with the PBFT, G-PBFT [19] and CPBFT [20]. When the number of nodes reaches 1000, the transaction delay in our scheme is about 27% of PBFT, 40% of G- PBFT and 31% of CPBFT. In throughput, our solution is 2.67 times larger than that of PBFT, 1.75 times larger than that of G-PBFT and 2.13 times larger than that of CPBFT. In terms of communication overhead, our scheme reduces by 75% compared with PBFT and 60% compared with CPBFT.

The rest of the paper is organized as follows: Section II introduces the related work on the identity authentication in the Internet of Vehicles and the improved PBFT. Sec- tion III introduces the PBFT consensus algorithm. Section IV introduces our scheme, SG-PBFT. Section V introduces our proposed IoV identity authentication model. Section VI analyzes communication overhead and safety of our scheme. Section VII presents our empirical analysis. Section VIII introduces the potential application of our scheme in the IoT. Section IX summarizes this paper and introduces the future work.

## II. RELATED WORK

In this section, we introduce the related work from two aspects: (1) identity authentication scheme in the IoV; (2) the improved PBFT. The identity authentication scheme in the IoV mainly includes the scheme of central server architecture and the scheme based on blockchain.

### A. Abbreviations and Acronyms

The basic principle of the scheme based on the central server is mainly based on the cryptography principle to ensure the security of users and vehicles. Wazid *et al*. [21] proposed a key management protocol for identity authentication based on fog computing. This protocol can be applied to the IoV systems and is called AKM-IoV. The scheme aims at achieving secure communication between vehicles by establishing session keys between entities. Li *et al*. [22] proposed the CL-CPPA proto- col. This protocol can meet the high requirements for privacy and security in the IoV system. Xu *et al*. [2] proposed the SE-CLASA protocol. This protocol is to reduce the communica- tion overhead while ensuring the safety of the vehicles. This protocol solves the problem of high computational overhead by adopting the solution of certificateless aggregation. Walshe *et al*. [23] proposed a non-interactive zero-knowledge authen- tication protocol that does not require authentication data to be stored on the device. This scheme can save calculation cost. However, the solution still needs to further evaluate the data protection processes introduced when calculating these hashes. Li *et al*. [24] proposed a PPA protocol for mobile services, which can provide efficient and secure mobile services. Li *et al*. [25] proposed a CreditCoin announcement to solve the problem that user identity information is not easily forwarded reliably and lacks forwarding power. This solution is very efficient in intelligent transportation. Zhu *et al*. [26] adopted a learn-based method to verify the ownership of mobile devices and formed a RISKCOG system. This scheme can be used in the environment without network, and can effectively verify the user's identity information. Jo *et al*. [27] proposed a two- layer pseudo-identity authentication method. This method can ensure that data packets are hardly lost. Therefore, the privacy of the vehicles and the users can be protected when the vehicle density is large.

The above schemes are all traditional authentication schemes, which have the widespread problems that the central server performance is prone to create bottlenecks and single point attacks. If the central server is attacked, the entire network may be down, the consequences are unimaginable. This problem can be solved by blockchain technology.

Due to the distributed blockchain technology, with advan- tages such as data integrity, non-tampering and privacy pro-tection, many scholars have studied the identity authentication scheme based on blockchain. Yin et al. [28] proposed a scheme to implement the safety information exchange between participating vehicles in the MCS network. Specifically, this paper deals with the situation of repetitive task allocation through an incentive mechanism, and at the same time designs an algorithm to sort the windows of the vehicles to improve the effective collaboration of safety information exchange in the IoV. Dai et al. [29] present an in-depth survey of the blockchain of things and made recommendations for the BCoT architecture. Feng et al. [30] proposed the BPAS framework.

This framework can reduce the dependence on the online center and can realize the dynamic tracking and cancellation of malicious vehicles. So as to better protect the privacy of the vehicles. Dorri et al. [31] proposed an intelligent vehicle ecological architecture based on blockchain. The solution is equipped with on-board storage devices that allow users to decide whether or not to provide certain information to third parties. Therefore, the safety of users and vehicles can be better protected. Yang et al. [32] proposed to verify the information of surrounding vehicles based on Bayesian inference, and write the information of trusted vehicles into blockchain through RSUs. This authentication scheme based on blockchain uses PoW consensus algorithm. The efficiency of authentication and computing overhead need to be improved. Liu et al. [33] proposed a vehicle group authentication scheme based on blockchain. By grouping RSUs with higher trust values, the members of each group can jointly authenticate a vehicle.

Although many scholars choose to adopt blockchain technology to solve the problem of central server, many solu- tions fail to consider the efficiency of consensus algorithm in blockchain. Low consensus efficiency and high communication overhead are common problems in such schemes. Therefore, some scholars use the improved PBFT consensus algorithm to solve these problems. In the next part, we will introduce the related work of the improved PBFT consensus algorithm.

*B. The imporved PBFT*

The distributed architecture of the blockchain certainly has benefits, but due to some problems in the practical application of the PBFT consensus algorithm, many scholars have made improvements in various aspects. Due to the relatively large communication overhead of the PBFT consensus algorithm, it is not suitable for dynamic networks. For this reason, Zhang et al. [34] proposed a PBFT consensus algorithm based on a grouping mechanism, which can improve the efficiency of transaction consensus and shorten the consensus time. A B- M tree is composed of Merkle tree and AVL tree. The B-M tree can realize an efficient way of querying data. Lao et al. [19] proposed a location-based and scalable PBFT consensus algorithm. Because the fixed node has stronger computing power than the mobile node, and the possibility of becoming a malicious node is very small. Therefore, the algorithm reduces the consensus overhead and ensures the security of the system by selecting a fixed and trusted node as the consensus participant. Onireti et al. [35] reduced the number of duplicate nodes in the PBFT consensus scheme by defining feasible areas for wireless networks, thereby reducing system overhead and improving system efficiency. Based on the P2P trust computing model, Tong et al. [36] proposed a distributed model that can be arbitrarily extended. By combining the PBFT consensus algorithm with the P2P trust computing model, the fault tolerance of the system and the trustworthiness of the system are improved. Li et al. [37] proposed Extensible- PBFT (EPBFT). This method uses verifiable random functions to select consensus nodes, changes the consistency protocol, simplifies the checkpoint protocol and the view conversion protocol, so that the consensus algorithm can be used in dynamic networks and reduces communication overhead.

In comparison with the work mentioned above, our SG-PBFT consensus algorithm uses the blockchain technology to build a distributed architecture to improve the identity authentication efficiency and reduce the communication overhead in the IoV environment. It also helps solve the problem of the central server being vulnerable to attacks. As indicated by our experiment reported in Section VII, our SG-PBFT algorithm can reduce the consensus time and improve the throughput, and also guarantee the security of the system.

III. BACKGROUND: PBFT CONSENSUS ALGORITHM

Blockchain technology involves various consensus algorithms. Reaching consensus in a distributed system depends on reliable consensus algorithms. It can ensure that the latest blocks are accurately added to the blockchain [38], and the blockchain information stored by the nodes is consistent and non-forking, and can even resist malicious attacks [39]. We can divide the consensus algorithm of the blockchain into two categories. One is the consensus algorithm between trusted nodes, and the other is the consensus algorithm between untrusted nodes. The former mainly includes Paxos, Raft and their corresponding variant algorithms [17]. The latter mainly includes consensus algorithms suitable for public blockchains represented by algorithms such as PoW and PoS, and suitable for consortium blockchains or private blockchains represented by PBFT and its variant algorithms [39]. Since the PBFT consensus algorithm is suitable for consortium blockchains or private blockchains, the algorithm is more in line with the application range of the Internet of Vehicles. This section mainly introduces the traditional PBFT consensus algorithm.

The traditional PBFT is mainly divided into three stages. These three stages are the pre-prepare stage, prepare stage, and commit stage. The normal process of this algorithm is shown in Fig. 1 and the pseudo code of PBFT consensus algorithm is shown in ALGORITHM 1.

C represents the client node and 0 to 3 nodes represent the consensus nodes. In particular, 0 is the master node and 3 is the failed node. 1 and 2 are replica nodes. ALGORITHM 1 is the pseudo code of the PBFT consensus algorithm. In this algorithm, o represents the operation requested by the client, t represents the timestamp, c represents the client's own identification, m represents the client's request, i represents the node, v represents the number of the current domain, n is the sequence number of the request, d is the requested hash value, f is the number of Byzantine nodes, and σ represents the signature. The specific process of the consensus algorithm is as follows:

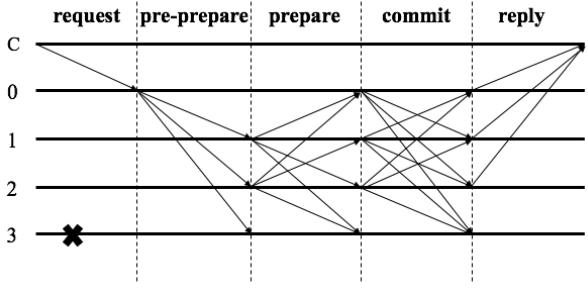

Fig. 1: Three-stage process of PBFT algorithm.

1. The client node initiates a request message to the master node.

2. After the master node has responded to the client request, it begins to enter the pre-prepare stage. The master node constructs a pre-prepared message and encodes the request message and broadcasts it to all replicated nodes.

3. The replica nodes receive and verify the pre-prepare message from the master node. If the validation is successful, the replica nodes will send the prepare message to the other consensus nodes.

4. The replica nodes will send the commit message to the other consensus nodes after the prepare message has been validated. When the request reaches committed state, it will be executed by the node.

5. For the same request, if the client node receives $2f + 1$ identical return results, it is considered that a consensus has been reached.

**ALGORITHM 1: PBFT**

```
whlie ⟨REQUEST, o, t, c⟩σ_c = TRUE do
    broadcast ⟨⟨PRE − PREPARE, v, n, d⟩σ_p, m⟩;
    if ⟨PRE − PREPARE⟩ = TRUE{
        broadcast ⟨PREPARE, v, n, d, i⟩σ_i;
        receive ⟨PREPARE, v, n, d, n⟩σ_n;
    }
    else do nothing;
    if prepared(m,v,n,i) = TRUE {
        broadcast ⟨COMMIT, v, n, i⟩σ_i;
        broadcast ⟨COMMIT, v, n, n⟩σ_n;
    }
    else do nothing;
    if committed-local(m,v,n,d,i) = TRUE{
        do ⟨REQUEST, o, t, c⟩σ_c;
        reply result to Client;
    }
    else do nothing;
end
```

## IV. SG-PBFT CONSENSUS ALGORITHM

We proposed the SG-PBFT consensus algorithm to improve consensus efficiency. The pseudo code of the SG-PBFT consensus algorithm is shown in ALGORITHM 2 and the main process of SG-PBFT after optimization is illustrated in Fig. 2. In this algorithm, o stands for the operation requested by the client, c represents the client's own identification, m represents the client's request, i represents the node, v represents the number of the current domain, n is the sequence number of the request, d is the requested hash value, f is the number of

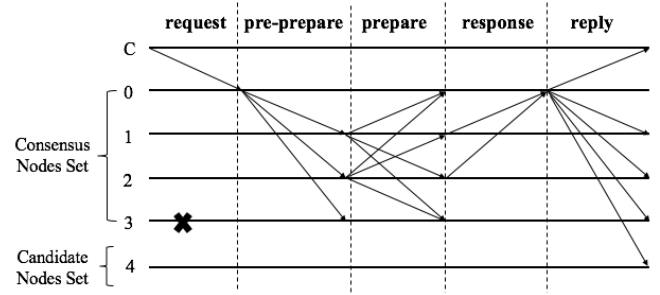

Fig. 2: Three-stage process of SG-PBFT algorithm.

**ALGORITHM 2: SG-PBFT**

```
whlie ⟨REQUEST, o, t, c⟩σ_c = TRUE do
    broadcast ⟨⟨PRE − PREPARE, v, n, d⟩σ_p, m⟩;
    if ⟨PRE − PREPARE⟩ = TRUE{
        broadcast ⟨PREPARE, v, n, d, i⟩σ_i;
        receive ⟨PREPARE, v, n, d, n⟩σ_n;
    }
    else do nothing;
    if prepared(m,v,n,i) = TRUE {
        result{}= result{} ⋃ result{i}σ_m
    }
    else do nothing;
    if ∃ ∑(result{i} ∈ result{}) > (2f + 1){
        reduce the trust value of Byzantine consensus
        nodes by 20;
        the master node will broadcast the result{},and
        the consensus nodes will update the result to their
        own state machines.
    }
    else do nothing;
    count_request++;
    if count_request=50{
        UpdateConNodes();
        count_request=0;
    }
end
```

Byzantine nodes, σ represents the signature, and t stands for the timestamp. It takes the following seven stages:

1. We initialize all nodes. Each node has a score of 100 scores and divides it into a consensus nodes set or a candidate nodes set. When the nodes are initialized, the order of all nodes is random. The consensus nodes set will execute the consensus process. And the candidate nodes only update their local states when reaching a consensus, but don't participate in the consensus process temporarily. There are N nodes in the system and the maximum number of Byzantine nodes is allowed to be f. We assume that there are N/2 candidate nodes and the number of consensus nodes must satisfy $CN \geqslant 3f+1$ to ensure that the system can reach an agreement in any case, where CN is the number of consensus nodes. So we let the number of consensus nodes be N/2.

2. We define the selection strategy of the master node as p =

v mod CN. The p represents the master node and v represents the view number.

3. The client node initiates a request message to the master node.

4. After the master node receives the request from the client node, it begins to enter the pre-prepare stage. The master node broadcasts the message to all replica nodes. It will verify the message, when the replica node receives the message. It will enter the prepare stage, if the verification passes. If the verification fails, then no action will take place.

5. If a node passes a pre-prepared message, the node enters the preparation stage. Then the node broadcasts the prepare message. When the replica node sends a prepare message, it will also receive the prepare message broadcast by other nodes and verify it. If it passes the verification, it will enter the response stage.

6. The node will verify the message and supply the result back to the master node, if a node enters the response stage.

7. It is considered to have reached a consensus, when the master node receives more than $2f + 1$ identical confirmation messages. The master node will then provide the confirmation result to all nodes (including candidate nodes), and updates the score of all nodes at the same time. For the node whose judgment result is consistent with the final result, the scores are increased by one. If the judgment result is inconsistent with the final result, the scores will be reduced by five. Nodes in the candidate and consensus sets are updated every 50 requests. The m nodes with the lowest score will be removed from the consensus set and instead will be appended to the end of the candidate list. The m nodes with the top scores in the candidate set will be appended to the consensus nodes set, and these nodes are renumbered. At this point, a consensus process is over.

## V. IDENTITY AUTHENTICATION SCHEME BASED ON SG-PBFT

The network model of our solution is shown in Fig. 3. The model includes three main entities: the vehicles, the roadside units (RSUs), and the service providers (SPs). The SP is mainly responsible for system parameters initialization and the vehicle registration. The vehicle submits its information to the SP. After that, the SP verifies its authenticity and completes the registration. The RSU fixed on the roadside uses blockchain technology to verify the identity of the vehicles. Finally, the RSU broadcasts the result to complete the verification of the vehicle.

We will explain the main contents of the scheme. This scheme mainly includes three parts: system initialization stage, vehicle registration stage, and certification stage.

First of all, the entire system needs to be initialized. Then, before sending and receiving messages, each vehicle needs to be registered. Both tasks are completed by the SP. Finally, the RSUs determine whether to write information about vehicles into the blockchain which is depended on the consensus result.

*1) Vehicle Initialization:* The SP initializes the parameters for the entire system. The initialization process has three phases:

a. The roadside unit (OBU) on the vehicle obtains the system parameters from the network. SP chooses two large prime numbers $p, q$ and a q-order group $G$. The generator of this group is the point $P$ on the nonsingular elliptic curve $E$. $E$ is a nonsingular elliptic curve $y^2 = x^3 + ax + b(mod\ p)$. ($a, b \in F_p$, $F_p$ is the range of the elliptic curve).

b. SP chooses a random number $P_{pri}$ as the system's private key and calculates its corresponding public key as $P_{pub}$.

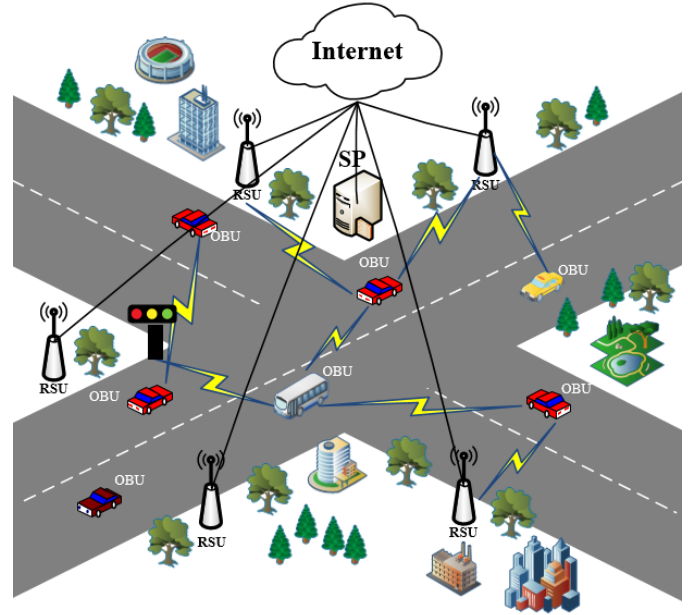

Fig. 3: The network model of IoV.

c. SP chooses two hash functions $H_1 : \{0,1\}^*_q \rightarrow Z^*_q$ and $H_2: G \rightarrow Z^*_q$. The common parameters of SP are $\{p,q,a,b,P, P_{pub}, H_1, H_2\}$.

*2) Vehicle Registration:* The vehicle encrypts all its information and sends it to the SP. The SP generates a random number and calculates the pseudo *ID* of a vehicle. After that, the SP encrypts it and sends it to the vehicle to complete the registration.

a. Each vehicle $V_i$ chooses an $ID_{Vi}$ as its vehicle's identity and then chooses a password $PW_{Vi}$.

b. The vehicle sends this information to the SP.

c. SP generates a random number $t_{i,j}$. The SP calculates $T_{i,j}$ as the registration time of the vehicle. Encrypt the pseudo *ID* to get $h_{i,j}$, and encrypt the private key to get $S_{i,j}$.

d. SP will send $(T_{i,j}, S_{i,j})$ to the OBUs and the user.

*3) Vehicle Authentication:* The RSU receives the vehicle registration message from the SP and verifies it. At this point, the SP is the client and the RSUs are the nodes on the blockchain. The RSUs verify it through the consensus algorithm in the blockchain to determine whether they can reach consensus. If the RSUs choose to trust the vehicle, the vehicle is granted a digital signature. Otherwise, they refuse to write the vehicle's information into the blockchain. However, the traditional consensus algorithm has the problems of low consensus efficiency and wasted computing power. Therefore, we propose the SG-PBFT consensus algorithm to improve authentication efficiency.

In Section VI and VII, we will compare the performance and efficiency of the traditional PBFT algorithm and the SG-PBFT algorithm proposed in this paper through experiments.

## VI. Theoretical Analysis

### A. Communication Overhead Analysis

Regarding the advantages of SG-PBFT consensus algorithm in communication overhead, we have made the following comparison calculations.

From Fig. 1, we can find that the specific calculation process of PBFT has three stages. In the pre-prepare stage, the number of communications in the consensus network at this stage is $(n-1)$ times. In the prepare stage, the number of communications in the consensus network at this stage is $(n-1)\cdot(n-1)$ times. In the commit stage, all nodes verify the received prepare message. When the verification result is true, the replica node sends a confirmation message to the other nodes except itself. The number of communications in the consensus network at this stage is $n\cdot(n-1)$. Therefore, the number of communications of the traditional PBFT consensus algorithm is $2n\cdot(n-1)$ within a consensus process.

Fig. 2 shows that the specific calculation process of the PBFT consensus algorithm has three stages. Let n denote the total number of nodes and the n/2 denote the number of consensus nodes. In the pre-prepare stage, the master node broadcasts the preparation message to all consensus nodes. The number of communications in the consensus network at this stage is $(n/2-1)$. In the prepare stage, the replica nodes send the prepare message to the other consensus nodes when the verification has passed the pre-prepare message sent by the master node. The number of communications in the consensus network at this stage is $(n/2-1)\cdot(n/2-1)$. In the response stage, the master node receives the verification message from the consensus nodes and verifies it. The number of communications in the consensus network at this stage is $(n/2-1)$. Therefore, the number of communications of the improved SG-PBFT consensus algorithm to complete a consensus process is $(n/2-1)\cdot(n/2+1)$. It can be inferred that the number of communications required to complete a consensus based on the G-PBFT [19] algorithm is about $(n+3)\cdot n/2$. And the CPBFT [20] consensus algorithm requires $n\cdot(n-1)$ times to complete a consensus.

So we can easily infer that the communication overhead of the SG-PBFT consensus algorithm is much lower than that of the traditional PBFT and CPBFT. Although our SG-PBFT consensus algorithm has almost the same as the communication overhead of the G-PBFT consensus algorithm, our solution is safer than G-PBFT. It is worth sacrificing a little communication overhead in exchange for system security, because security is more important in practical applications.

### B. Safety Analysis

The security of the blockchain architecture is crucial. Once the blockchain is attacked, it will lead to the leakage of user information, vehicle location information, and other private information. The security analysis of the system is as follows.

#### 1) Proof of PBFT

The logical proof of the SG-PBFT algorithm is as follows: We assume that N is the number of consensus nodes, $f$ is the number of Byzantine nodes, $M$ is the set of messages, and $n_i$ is the node i.

$$N = 4$$
$$f = 1$$
$$M = \{m_0, m_1, m_2, ..., m_i, ...\}$$

where $m_i$ represents a message, and i is an integer from 0 to $\infty$.

Let $M_{ni}$ is the messages received by $n_i$. So, $M_{ni}$ can be expressed by the following equation:

$$M_{ni} = \{m_0 n_i, m_1 n_i, ..., m_i n_i, ...\}$$

In the request stage, the client sends a message $M$ to the master node $n_0$. So, we can get the following equation:

$$M_{n0} = \{m_0 n_0, m_1 n_0, ..., m_i n_0, ...\}$$

In the pre-prepare stage, the master node $n_0$ broadcasts the message $M$ to all replica nodes. If the replica node $n_i$ is normal, then

$$M_{ni} = M_{n0}$$

Here, we define if

$$M_{ni} = M_{n0}$$

then,

$$\langle pre-prepared \rangle\, M_{ni} = 1$$

otherwise,

$$\langle pre-prepared \rangle\, M_{ni} = 0$$

where $\langle pre-prepared \rangle\, M_{ni}$ is the verification of message $M$ by node $n_i$. We know that

$$N = 4$$
$$f = 1$$
$$N \geq 3f + 1$$

So, we can get:

$$\sum_{i=1}^{3} \langle pre-prepared \rangle\, M_{ni} \geq 2f$$

In the prepare stage, if

$$\langle pre-prepared \rangle\, M_{ni} = 1$$

then,

$$\langle prepared \rangle\, M_{ni} = 1$$

otherwise,

$$\langle prepared \rangle\, M_{ni} = 0$$

where $\langle pre-prepared \rangle\, M_{ni}$ is the verification of the pre-prepare message by node $n_i$.

In this stage, for any node participating in the consensus, it will also receive prepare messages sent by other nodes when it sends a prepare message. Then for node ni, there are the following relations, where $\langle prepared \rangle n_i$ is the number of correct prepare messages:

$\langle prepared \rangle n_0 = \langle prepared \rangle M_{n1} + \langle prepared \rangle M_{n2} + \langle prepared \rangle M_{n3}$

$\langle prepared \rangle n_1 = \langle prepared \rangle M_{n0} + \langle prepared \rangle M_{n2} + \langle prepared \rangle M_{n3}$

$\langle prepared \rangle n_2 = \langle prepared \rangle M_{n0} + \langle prepared \rangle M_{n1} + \langle prepared \rangle M_{n3}$

$\langle prepared \rangle n_3 = \langle prepared \rangle M_{n0} + \langle prepared \rangle M_{n1} + \langle prepared \rangle M_{n2}$

As we know, the following equation is always established:

$$\langle prepared \rangle\, M_{n0} = \langle prepared \rangle\, M_{n1} = \langle prepared \rangle\, M_{n2} = 1$$

so, we can get this result:

$$\langle prepared \rangle n_i \geq 2f\ (i = 0, 1, 2, 3)$$

In the response stage, according to $\langle prepared \rangle n_i \geq 2f$, at least nodes $n_0$, $n_1$ and $n_2$ can process the message $M$ correctly, so the master node $n_0$ will receive message $M_{ni}$ satisfies the following formula:

$$\langle response \rangle ni \geq 2f + 1$$

We can believe that it is safe to reach a consensus.

According to the above proof, we can know that our scheme can guarantee the reliability of nodes, the correctness of each round of results, prevent tampering and forgery, and resist DoS attacks.

*2) Credibility of nodes*

When the master node is a Byzantine node, the master node is either likely to broadcast an inconsistent pre-prepare message to all consensus nodes or not respond to client requests. Because if this abnormal situation occurs, it will cause the prepare stage to reach a consensus failure and thus initiate the view change to replace the master node. If the current node is a malicious master, our scheme will reduce the trust of the master node. In the subsequent operation of updating the set of consensus nodes, the malicious master node is moved to the set of candidate nodes. Therefore, the credibility of the nodes in the consensus nodes set can be guaranteed.

When the number of Byzantine nodes in the consensus node set is f, since the total number of the consensus node sets is more than or equal to $3f + 1$, any node in the prepare and response stages can always receive more than or equal to 2f consistent messages because it will not affect the process of reaching consensus. In summary, the SG-PBFT consensus algorithm can verify every round, and the results are credible.

*3) Prevention of Tampering and Forgery*

In our scheme, the vehicle uses pseudo identification, hash algorithm and digital signature. Moreover, the blockchain uses asymmetric encryption technology and SG-PBFT consensus algorithm to prevent data tampering. Therefore, any node without the user's private key cannot forge the identity of the node and the signature of the message. By using blockchain technology and digital signature, the forgery cost of attackers is too high, thus ensuring the reliability of the whole system.

*4) Resist DDos Attack*

In order to resist DDoS attack, we use score mechanism to select the master node. The identity of the master node will not be exposed before the proposal. Because after each consensus, we adjust the nodes in the consensus node set and the candidate node set, and renumber them. This can ensure that the identity of the master node is hidden, and the enemy can not get the master node information. The identity of the master node is not exposed until recommended. Because after each consistency, we adjust and renumber the nodes in the consistency node set and the nodes in the candidate node set. This can ensure that the identity of the master node is hidden, and the enemy can not get the information of the master node. Therefore, the enemy can not send a large number of invalid information to the master node to occupy the system resources. Therefore, our scheme can effectively resist DDoS attack and ensure the security of the system.

## VII. Empirical Analysis

In this paper, we simulate the performance of PBFT, SG-PBFT, G-PBFT [19] and CPBFT [20] consensus algorithms by C++ language. In this experiment, we regard vehicles as nodes in the blockchain and simulate the situation of different numbers of vehicles in the IoV by setting different numbers of nodes. If the program can output the correct transaction results, we believe that a consensus has been reached. The consensus reached by these nodes means that the vehicles in the IoV have reached a consensus. We set the initial score of the node to 100, and determine the score value of each node after each consensus through the above algorithm. In our experiments, we use a PC(ThinkPad with an i7-5500U 2.4GHz processor,8GB memory) to test the performance of the four consensus algorithms. We test and analyze the four aspects: transaction delay, throughput, communication overhead, and the safety of the four consensus algorithms.

### A. Transaction delay

First, we test the transaction time delay of the four algo- rithms. The transaction time delay refers to the time taken from the moment when the client node sends a transaction request to the master node to the moment when the client receives the transaction confirmation message. We conducted two hundred tests and use the average value of every ten transaction delays as the experimental data, thus guaranteeing the generality of the experimental results. Fig. 4 shows the transaction time delay of the four consensus algorithms. We can find from Fig. 4, the transaction time delay of SG-PBFT is significantly lower than that of PBFT, G-PBFT [19] and CPBFT [20]. But the result shows that the SG-PBFT has obvious advantages over the other three methods in terms of transaction delay.

The main reason for the above results is that our scheme reduces the amount of information sent to each other and reduces the number of nodes participating in consensus. In our scheme, we use the response and reply stages to ensure that each node can be in a consistent state. By replacing the commit stage, the transaction delay is reduced.

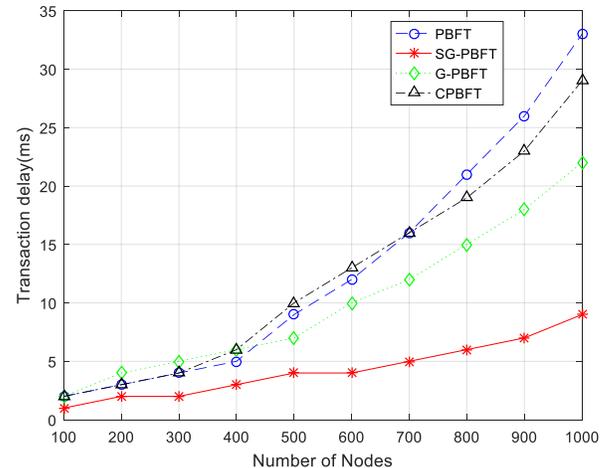

Fig. 4: The transaction time delay of the four consensus algorithms.

### B. Throughput

In the second part of the experiment, we test the through- put of the transaction. The throughput is the number of transactions completed by the system per unit time. In this experiment, we set up the client to send 3000 requests and record the number of transactions completed per second with different numbers of nodes. We use this method to test the throughput of the transaction. Fig. 5 is the comparison chart of the throughput of the four consensus algorithms. Fig. 5 shows that the throughput of the SG-PBFT consensus algorithm is higher than that of the traditional PBFT, G-PBFT [19] and CPBFT [20] consensus algorithms. At the same time, with the increase of the number of nodes, the throughput of the four algorithms all showed a downward trend. The advantages of SG-PBFT algorithm are still very obvious.

The main reason for this phenomenon is that the delay of each transaction is smaller than that of PBFT, G-PBFT [19] and CPBFT [20] consensus algorithms, as shown in A. Therefore, in a unit of time, our scheme can complete more transactions, and the throughput is higher than other schemes.

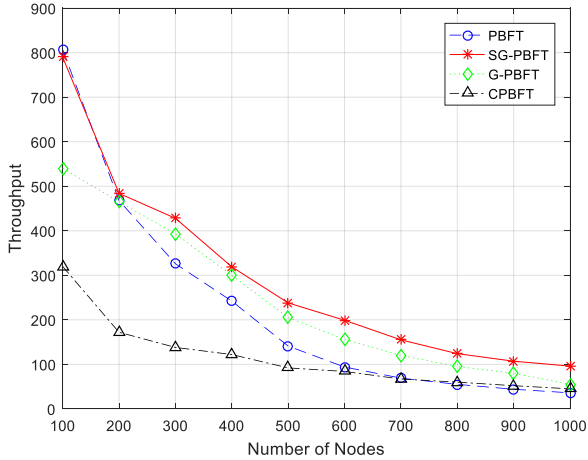

Fig. 5: The throughput of the four consensus algorithms.

### C. Communication overhead

In the third part of the experiment, we test the communication overhead of the transaction. Communication overhead refers to the amount of the communication generated when the nodes in the system execute the consensus algorithm. In this experiment, we tested the single transaction traffic of the four consensus algorithms separately. Fig. 6 is a comparison chart of the traffic of a single transaction between four consensus algorithms. Fig. 6 shows that with the increasing number of the consensus nodes, the communication overhead of the SG-PBFT and G-PBFT [19] consensus algorithm grows slowly. But the communication overhead of the traditional PBFT and CPBFT [20] consensus algorithm increases rapidly.

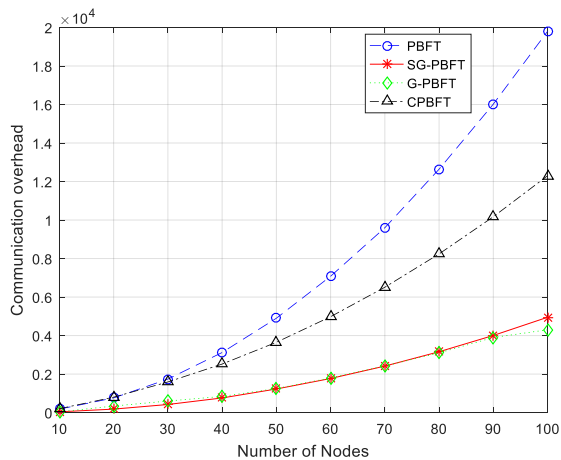

Fig. 6: The communication overhead of the four consensus algorithms.

## VIII. PITENTIAL BENFICIAL IOT APPLICATIONS

Our distributed architecture is not only suitable for the IoV system but also suitable for other industrial Internet of Things applications [40], such as the smart grid, the intelligent manufacturing and so on. Now we take the smart grid as an example. Fig 7 is the smart grid structure. It is composed of a central server, regional gateways, and clients(smart meters). Obviously, this is a C/S architecture. As we know, once a server with a C/S architecture is subjected to a malicious attack and causes a crash or tampering, the entire system will be in great danger. In the smart grid, the wireless characteristics of the smart grid equipment are easy to cause user privacy leakage [41]. In addition, as the power demand of smart grid users increases, a large number of smart grid service requests and responses may cause network congestion problems, further increasing the risk of server crashes [42]. If the central server or regional gateway of the smart grid is attacked, this may result in leakage or tampering of user information. It may even cause the entire system to enter a paralyzed state. This has very serious consequences. At this time, malicious attackers may launch simulated attacks to obtain free services, and may even endanger users' money and lives.

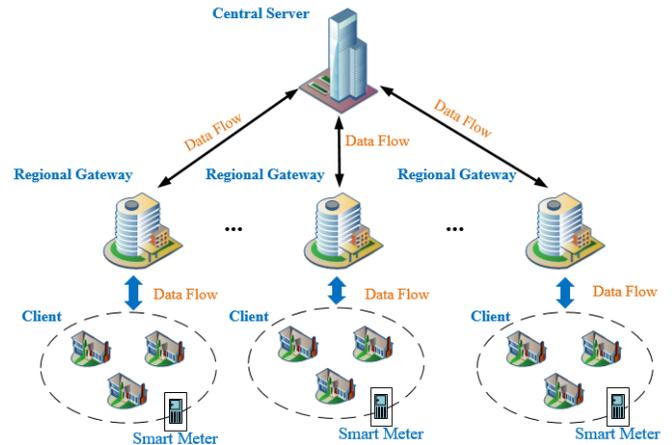

Fig. 7: The Smart Grid Model.

The identity authentication scheme based on blockchain proposed by us can be applied to smart grid [43]. Specifically, as shown in Fig 8, the central server and regional gateway can be eliminated. We can give the smart meters of computing power through hardware technology. We regard smart meters as blockchain nodes. Each smart meter registers identity information with the service provider and obtains a private key. When a new smart meter joins the blockchain, other smart meters already on the chain should authenticate the user information of the smart meter that is about to join the blockchain. At the same time, we can use our proposed SG-PBFT consensus algorithm to improve the efficiency of the user authentication.

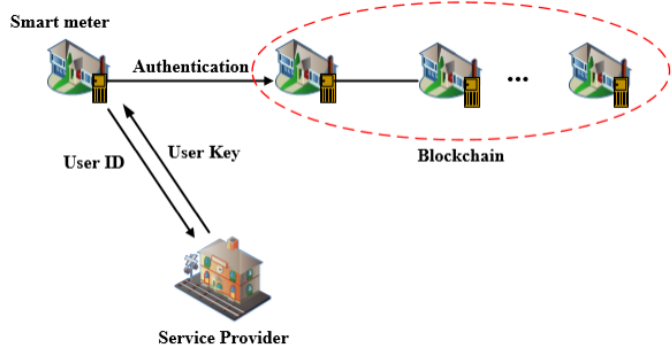

Fig. 8: Smart grid model based on blockchain.

## IX. Conclusion

In this paper, we proposed a novel PBFT, called SG-PBFT, based on a score grouping system. The SG-PBFT consensus algorithm not only considers the efficiency of the consensus, but also considers the security of the system. We also divide the nodes into groups according to the score, which further improved the system consensus efficiency. The experimental results show that SG-PBFT is superior to PBFT, G-PBFT and C-PBFT in transaction delay, throughput and communication overhead. At the same time, we use the integrity, traceability, non tampering and high transparency of the blockchain to apply the scheme to the identity authentica- tion of the IoV system. We use Vehicle - Service Provider - Roadside Unit architecture to ensure the security of the whole system. In the future work, we will study how to optimize the node selection strategy to further shorten the consensus time and improve the consensus efficiency. On this basis, further attempts to study the use of blockchain technology to solve the network security problems caused by malicious vehicles in the IoV system, so as to provide more secure and efficient services for the IoV system.

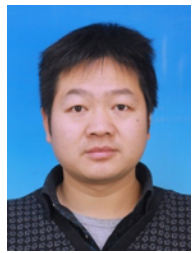

**Guangquan Xu** is a Ph.D. and full professor at the Tianjin Key Laboratory of Advanced Networking (TANK), College of Intelligence and Computing, Tianjin University, China. He received his Ph.D. degree from Tianjin University in March 2008. His research interests include cyber security and trust management. He is the director of Network Security Joint Lab and the Network Attack Defense Joint Lab. He has published 100+ papers in reputable international journals and conferences, including IEEE Transactions on Cybernetics, IEEE Internet of Things Journal, ACM Transactions on Internet Technology, ACM Transactions on Intelligent Systems and Technology, IEEE Transactions on Industrial Informatics, Information Sciences, IEEE Wireless Communications, IEEE Network, Computers Security, and so on. He served as a TPC member for IEEE UIC 2018, SPNCE2019, IEEE UIC2015, IEEE ICECCS 2014 and reviewers for journals such as IEEE access, ACM TIST, JPDC, IEEE TITS, soft computing, FGCS, and Computational Intelligence, and so on. He is a Fellow of IET, IEEE member, senior member of China Computer Society. Email:losin@tju.edu.cn(https://orcid.org/0000-0001-8701-3944)

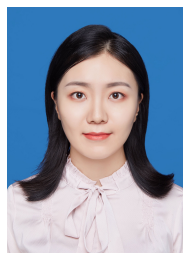

**Yihua Liu** is a master's student at the International Engineering Institute, Tianjin University, China. She received her B.S. degree from the School of Communication Engineering, Hebei University of Technology of China in 2018. Her current research interests include blockchain and Internet of Vehicle. Email: 13920878179@163.com

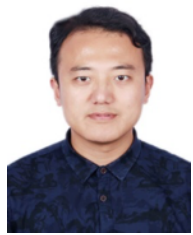

**Jun Xing** is a lecturer of Big Data School, Qingdao Huanghai University. He received the Ph. D. degree from Naval University of Engineering in 2009. He is a member of the CCF. His research interests include application of big data technology, Cognitive intelligence, and Data analysis and mining. Email:fumulan cn@163.com

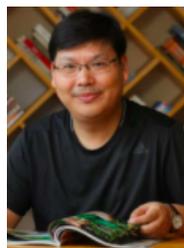

**Tao Luo** received the M.S. degree from the School of Precision Instrument and Opto-Electronics, in 2006, and the Ph.D. degree from the School of Electronic Information Engineering, Tianjin University, in 2009. He is currently an Associate Professor with the College of Intelligence and Computing, Tianjin University. His research interests include Internet security; intelligent vision processing, and integrated circuits. Email: luo tao@tju.edu.cn.

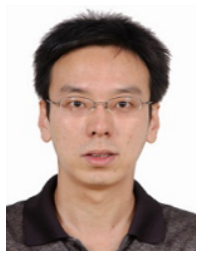
**Yonghao Gu** received his Ph.D. (2007) from Beijing University of Posts and Telecommunications, China. Currently, he is a master supervisor in the Beijing Key Laboratory of Intelligent Telecommunications Software and Multimedia, School of Com- puter, Beijing University of Posts and Telecommunications, China. His main research interests are network security and big data security. E-mail:guyonghao@bupt.edu.cn.

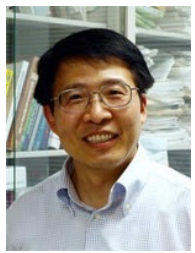
**Shaoying Liu** holds a B.Sc and a M.Sc degree in Computer Science from Xi'an Jiaotong University, China, and the Ph.D in Computer Science from the University of Manchester, U.K. He worked as Assistant Lecturer and then Lecturer at Xi'an Jiaotong University, Research Associate at the University of York, and Research Assistant in the Royal Holloway and Bedford New College at the University of London, respectively, in the period of 1982 -1994. He joined the Department of Computer Science at Hiroshima City University as Associate Professor in April 1994, and the Department of Computer Science in the Faculty of Computer and Information Sciences at Hosei University in April 2000. In April 2001 he was promoted to a Professor. From 1st April 2020, he has been working at Hiroshima University as a Professor. He was invited as a Visiting Research Fellow to The Queen's University of Belfast from December 1994 to February 1995, a Visiting Professor to the Computing Laboratory at the University of Oxford from December 1998 to February 1999, and a Visiting Professor to the Department of Computer Science at the University of York from April 2005 to March 2006. From 2003 he is also invited as an Adjunct Professor to Shanghai Jiaotong University, Xi'an Jiaotong University, Xidian University, and a Visiting Professor to Shanghai University, Xi'an Polytechnic University, Bejing Jiaotong University, and Beijing University in China, respectively. He is IEEE Fellow, British Computer Society (BCS) Fellow, and member of Japan Society for Software Science and Technology. Email:sliu@hiroshima-u.ac.jp

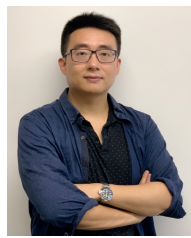
**Xi Zheng** got PhD in Software Engineering from UT Austin. He specialised in Machine Learning Testing, Distributed learning and Embedded Intelligence, IoT Security and Reliability Analysis. Now Director of Intelligent Systems Research Group and Associate Professor/Senior Lecturer in Software Engineering at Macquarie University. Published more than 80 high quality publications in top journals and conference. PC for PerCom and TrustCom. Awarded the best paper in Australian distributed computing and doc- toral conference in 2017. Awarded Deakin Research outstanding award in 2016 and Macquarie Early Career Research Highly Commended in 2020. Awarded Multiple ARC LP and DP projects. Active reviewer for top journals and conferences. Email: james.zheng@mq.edu.au

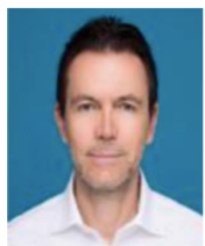
**Athanasios V. Vasilakos** is with the University of Technology, Sydney, Australia, with the Fuzhou University, Fuzhou, China and Lulea University of Technology, Lulea, Sweden. He has authored or coauthored 600 papers in peer-reviewed journals and conferences, with 36 000 citations. His main research interests include cybersecurity, networking, the IoTs and big data analytics. Prof. Vasilakos is an Editor for many technical journals, such as the IEEE Transactions on Network and Service Management, the IEEE Transactions on Cloud Computing, the IEEE Transactions on Information Forensics and Security, the IEEE Trans- actions on Cybernetics, the IEEE Transactions on NanoBioscience, and the ACM Transactions on Autonomous and Adaptive Systems. He is also the General Chair of the European Alliances for Innovation.